\documentclass[journal,comsoc]{IEEEtran}
\usepackage[T1]{fontenc}

\usepackage{cite}

\usepackage[pdftex]{graphicx}
\graphicspath{{Figures/}}

\usepackage{amsmath}
\usepackage[cmintegrals]{newtxmath}
\usepackage{algorithmic}
\usepackage{algorithm}
\usepackage[caption=false,font=normalsize,labelfont=sf,textfont=sf]{subfig}

\usepackage[dvipsnames]{xcolor}
\begin{document}
%
\title{Gradient-free training of autoencoders for non-differentiable communication channels}
%
%
%

\author{Ognjen~Jovanovic,~\IEEEmembership{Student Member,~IEEE,}
        Metodi~P.~Yankov,~\IEEEmembership{Member,~IEEE,}
        Francesco~Da~Ros,~\IEEEmembership{Senior Member,~IEEE,}
        and~Darko~Zibar
\thanks{O. Jovanovic, M. P. Yankov, F. Da Ros, and D. Zibar are with the Department
of Photonic Engineering, Technical University of Denmark, 2800 Kgs. Lyngby,
Denmark, e-mail: ognjo@fotonik.dtu.dk}
}

%
%

\markboth{JLT DRAFT}%
{Preprint}
%



\maketitle

\begin{abstract}
Training of autoencoders using the back-propagation algorithm is challenging for non-differential channel models or in an experimental environment where gradients cannot be computed. In this paper, we study a gradient--free training method based on the cubature Kalman filter. To numerically validate the method, the autoencoder is employed to perform geometric constellation shaping on differentiable communication channels, showing the same performance as the back-propagation algorithm. Further investigation is done on a non--differentiable communication channel that includes: laser phase noise, additive white Gaussian noise and blind phase search-based phase noise compensation. Our results indicate that the autoencoder can be successfully optimized using the proposed training method to achieve better robustness to residual phase noise with respect to standard constellation schemes such as Quadrature Amplitude Modulation and Iterative Polar Modulation for the considered conditions.     
\end{abstract}

\begin{IEEEkeywords}
Optical fiber communication, cubature Kalman filter, end-to-end learning, geometric constellation shaping, phase noise.
\end{IEEEkeywords}

%
\IEEEpeerreviewmaketitle

\section{Introduction}
\IEEEPARstart{C}{ommunication} systems consist of a transmitter and receiver designed with the aim to reliably transfer information from one end to another over a physical channel, e.g. air or optical fiber. Typically, both the transmitter and receiver are structured as a chain of multiple independent signal processing blocks, such as channel coding, modulation, pulse shaping, and equalization. Even though this block-wise approach has proven to be efficient, it is uncertain if it achieves the best possible end-to-end performance, e.g. highest throughput.

End-to-end learning of a communication system, which optimizes the transmitter and receiver (including all of their processes) jointly for a specific channel model and performance metric was introduced in \cite{OShea2017}. A communication system is perceived as an autoencoder (AE) \cite{goodfellow2016deep} by representing the transmitter and receiver as neural networks (NNs) and considering the channel model as a non-trainable layer. Optimizing the AE is typically done by applying gradient--based algorithms, which require a \emph{differentiable} channel model. In \cite{Dorner}, the AE was trained numerically for a wireless link and its benefits were demonstrated experimentally. The idea has been expanded and applied to orthogonal frequency-division multiplexing (OFDM) \cite{Felix} and multiple--input multiple--output (MIMO) \cite{MIMO}. In order to optimize full waveforms for optical fiber transmission, the AE has been applied to a dispersive linear fiber channel \cite{Karanov2018,Karanov2019a}, and nonlinear frequency division multiplexing (NFDM) \cite{Gaiarin2020,GaiarinJLT} for transmission over the nonlinear dispersive fiber channel model.

The AE approach can be utilized for geometric constellation shaping (GCS) in order to close the gap to the theoretically achievable information rate, which is experienced with uniformly distributed signalling (such as conventional quadrature amplitude modulation (QAM)). Geometric constellation shaping is a process of optimizing the positions of the constellation points on the I/Q plane. The main goal of GCS is to achieve the best possible trade--off between Euclidean distance and energy distribution of the constellation for the given channel. Embedding the AE with different optical fiber channel models to learn geometric constellation shapes was demonstrated for perturbative models of the nonlinear fiber channels \cite{Jones2018a,Jones2019}, for non-dispersive channels \cite{Li2018b} or for linear links with up to 1.2 dB of optical signal-to-noise ratio (SNR) gain with respect to standard QAM \cite{Schaedler2020}.
 
All the aforementioned works fulfilled the requirement that the channel model is differentiable in order to perform classical gradient--based optimization of an AE. This requirement is often too strict because 1) not all channel models are differentiable; 2) approximating a channel with a simple differentiable model results in inaccuracies \cite{Karanov2020}; 3) an accurate differentiable channel model can be too complicated for reliable optimization.

As an alternative, it was proposed to use generative adversarial networks (GANs) to provide a simple differentiable channel model of the complex or non-differentiable observed system \cite{o2019approximating}. The GANs are used to train the AE as demonstrated in \cite{ye2018channel} for wireless communication and \cite{Karanov2020} for non-coherent fiber optic communication. However, using a GAN as a channel model adds an extra step to the AE optimization. For every optimization step of the AE, new data needs to be obtained from the non-differentiable channel model and used to train the GAN. The GAN requires a lot of training samples to learn an accurate channel approximation. Therefore, the overall process of AE optimization using GANs can be time-consuming \cite{Karanov2020}. Moreover, the data generated by the GAN during the AE optimization is synthetic and it is just an approximation of the probability distribution of the actual channel. 

In \cite{Hoydis,Model-free}, a two--phase alternating algorithm for training of the AE without a channel model was presented. The algorithm is based on observing the transmitter and receiver as two separate NNs, applying a variant of reinforcement learning to train the transmitter and supervised learning to train the receiver. This approach is in contrast with a joint optimization since the optimization of the transmitter and receiver is not done simultaneously, and it requires more samples to converge \cite{Raj}. A simultaneous optimization approach for AE without channel knowledge using simultaneous perturbation stochastic approximation (SPSA) \cite{spall1992multivariate} for gradient estimation was demonstrated in \cite{Raj}. However, it was shown that the variance of the gradient estimation increases with the number of parameters \cite{Model-free}. As a consequence, SPSA struggles to train AEs with a large number of parameters, e.g. a large constellation size if targeting constellation shaping.

In this paper, the differentiable channel model requirement is lifted by proposing a derivative-free optimization method for AEs. This allows the encoder and the decoder to be optimized simultaneously for \emph{arbitrary black-box channels}, including non-numerical ones (e.g. experimental test-beds). Also, the method has the potential for online optimization. The proposed method is exemplified by adopting the cubature Kalman filter (CKF) \cite{haykin2009cubature} for the optimization of an AE. In order to show that this method can be used for differentiable channel models, the AE performs GCS for a differentiable AWGN and nonlinear phase noise channels, resulting in nearly identical performance to typically used gradient--based optimization. Then, the AE is trained to perform GCS on a phase noise channel with residual phase noise, resulting from a non--differentiable carrier phase recovery algorithm.

The remainder of the paper is organized as follows. In Section \ref{FGCS} the basic principles of estimating the mutual information and the key concepts of using an AE for GCS are described. Section \ref{E2E-AE} explains how the weights of the AE can be optimized using CKF. A detailed description of the system, the AE architecture and channel models, is provided in Section \ref{system}. Section \ref{results} provides results on the mutual information achieved by the AE for different channel models. In Section \ref{Discussion} the simulation results and future work are discussed. The conclusions are summarized in Section \ref{Conclusion}.

\emph{Notations:} Boldface denotes multivariate quantities such as vectors (lowercase) and matrices (uppercase). The sets of real and complex numbers are denoted as $\mathbb{R}$ and $\mathbb{C}$, respectively. The subscript $k$ and $j$ indicate time and iteration step, respectively. The covariance matrix of matrix $\mathbf{A}$ is denoted as $\mathbf{P}_{\mathbf{AA}}$, whereas the cross--covariance matrix of matrices $\mathbf{A}$ and $\mathbf{B}$ is denoted as $\mathbf{P}_{\mathbf{AB}}$. The $i$-th column of a matrix $\mathbf{A}$ is represented by $\mathbf{A}_i$, whereas the $i$-th scalar element of a vector $\mathbf{a}$ is represented by $\mathbf{a}^{(i)}$. The subscript $j|j-1$ is used to indicated that the current value of the matrix (vector) is conditional to the previous iteration. If two subscripts occur for the same matrix they are separated by a comma, e.g. $\mathbf{A}_{i,j|j-1}$ or  $\mathbf{P}_{\mathbf{AA},j|j-1}$.

\section{Fundamentals of Geometric constellation shaping}\label{FGCS}
\subsection{Mutual information}
Consider $X$ to be a sequence of complex constellation points (symbols) that take values from $\mathcal{X}=\{x_1,x_2,\dots,x_M\}$ with a uniform probability mass function $P_{X}(x)=\frac{1}{M}$, where $M$ is the number of constellation points. The entropy of the constellation $H(X)= -\sum_{x \in \mathcal{X}}P_{X}(x)\log_2(P_{X}(x))=\log_2(M)=m$ represents the number of bits carried by a symbol. Let $Y$ be a continuous complex output of a memoryless channel with $X$ as its input. The input-output relation of the channel is governed by the channel transition probability density $p_{Y|X}(y|x)$. The amount of information that $Y$ contains about $X$ in bits per symbol is represented by mutual information (MI)
\begin{equation}
\begin{split}
    I(X;Y) & = H(X)-H_p(X|Y) = m - H_p(X|Y) \\
    & = \sum_{x \in \mathcal{X}}P_{X}(x) \int_{\mathbb{C}} p_{Y|X}(y|x)\log_2 \frac{p_{Y|X}(y|x)}{p_{Y}(y)}dy \text{,} 
\end{split}
\label{eq:MI}
\end{equation}
where $\mathbb{C}$ denotes the set of complex numbers, $H_p(X|Y)= \mathbb{E}[p_{X|Y}(x|y)]$ is the conditional entropy of $X$ given $Y$ and $p_{Y}(y)$ is the probability density distribution of $Y$.

In order to calculate Eq. (\ref{eq:MI}), the transition probability $p_{Y|X}(y|x)$ must be known. Since the main topic of this paper is black-box channels (which do not have known explicit analytical expressions), the transition probability is unknown. In such cases, Eq. (\ref{eq:MI}) needs to be bound. A lower bound on the MI, also known as the achievable information rate (AIR), can be obtained by using the mismatched decoding approach and assume the transition probability $q_{Y|X}(y|x)$ of an auxiliary channel instead of the true $p_{Y|X}(y|x)$ \cite{Lowerbound},
\begin{equation}
\begin{split}
    I(X;Y) & \geq H(X)-\hat{H}_q(X|Y) = m - \hat{H}_q(X|Y) \\
    & = \sum_{x \in \mathcal{X}}P_{X}(x) \int_{\mathbb{C}} p_{Y|X}(y|x)\log_2 \frac{q_{Y|X}(y|x)}{q_{Y}(y)}dy \text{,} 
\end{split}
\label{eq:MI_lower}
\end{equation}
where $\hat{H}_q(X|Y)=\mathbb{E}[q_{X|Y}(x|y)]$ is the upper bound of the true conditional entropy $H_p(X|Y)$. The inequality turns to equality only when $q_{Y|X}(y|x)=p_{Y|X}(y|x)$.

\subsection{Geometric constellation shaping with autoencoders}
Typically, GCS involves the optimization of the constellation points in the complex plane with the aim to maximize the MI $I(X;Y)$. The MI can be maximized without explicit channel knowledge or assumption by leveraging the AEs to perform GCS. The AE accomplishes this by making an approximation $\hat{p}_{X|Y}(x|y)$ of the true posterior distribution $p_{X|Y}(x|y)$ using the decoder NN. It was shown that an AE can be used to improve the lower bound of the MI \cite{Li2018b}.

\begin{figure*}[!t] 
\centering
\includegraphics[width=\textwidth]{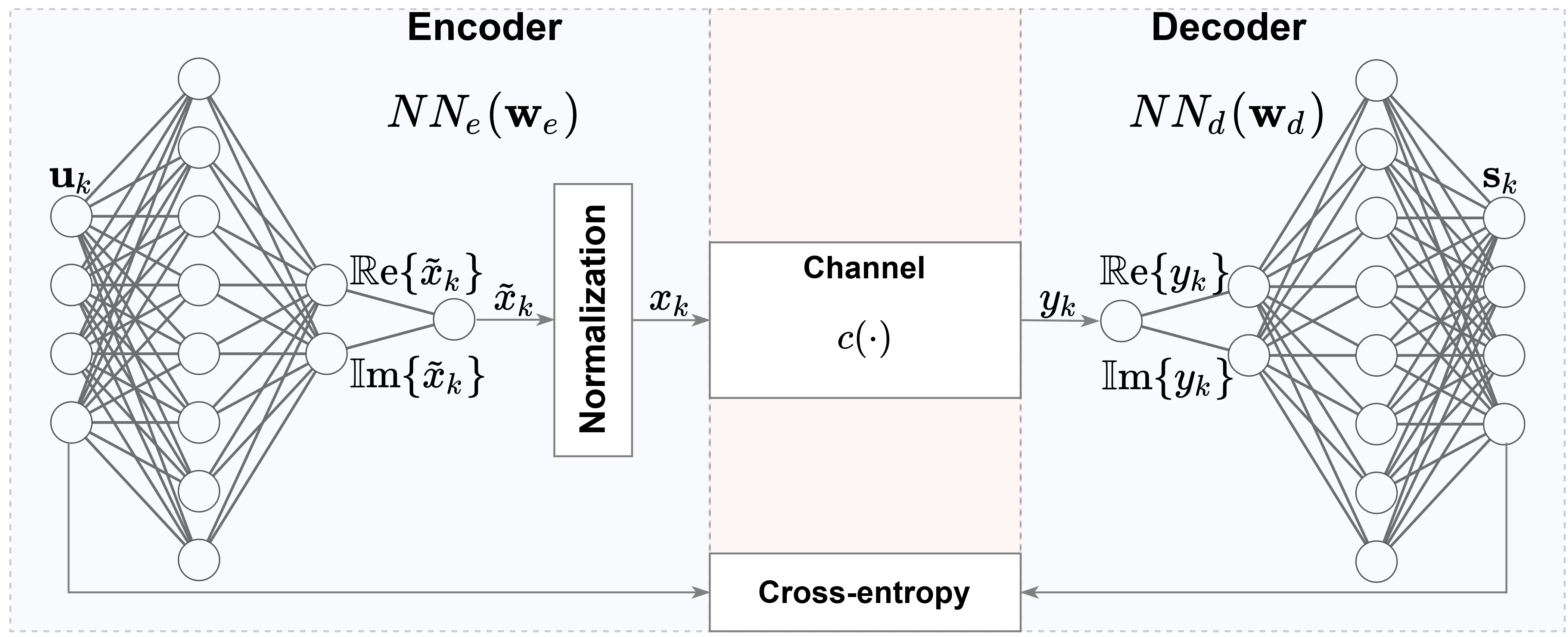}
\caption{Example of autoencoder model for geometrical constellation shaping. The number of input/output nodes of the AE and the hidden layers are used for illustration purposes.}
\label{fig:AE}
\end{figure*}

The considered communication system employing an AE for geometrical constellation shaping is shown in Fig. \ref{fig:AE}. The encoder and the decoder are represented by feed-forward neural networks $NN_{e}(\mathbf{w}_{e})$ and $NN_{d}(\mathbf{w}_{d})$, parameterized with trainable weights $\mathbf{w}_{e}\in \mathbb{R}^{N_e}$ and $\mathbf{w}_{d} \in \mathbb{R}^{N_d}$, respectively. The number of weights, including all trainable weights and biases of the NN, in the encoder is denoted by $N_e$, whereas the number of weights in the decoder is $N_d$. The overall goal is to find the corresponding $NN_{e}(\mathbf{w}_{e})$ and $NN_{d}(\mathbf{w}_{d})$ topologies and the weight set, $\mathbf{w}=\{\mathbf{w}_e,\mathbf{w}_d\} \in \mathbb{R}^{N}$, where $N=N_e+N_d$ is the total number of trainable weights, that would maximize the MI between the transmitted and the received symbols for the considered channel. In this paper, the goal of the encoder is to shape the input sequence to a geometrically optimized constellation that is robust to channel impairments, whereas the decoder learns to reconstruct the transmitted symbols with high fidelity.

The encoder $NN_{e}(\mathbf{w}_{e})$ performs a mapping of the input one--hot encoded vector $\mathbf{u}_k\in \mathbb{R}^M$ to an output $\tilde{x}_k = \operatorname{\mathbb{R}e}\{\tilde{x}_k\}+i \cdot \operatorname{\mathbb{I}m}\{\tilde{x}_k\} \in \mathbb{C}$, representing a point in a constellation plane, where $k$ is a running index representing time, $\operatorname{\mathbb{R}e}\{\cdot\}$ is the real part and $\operatorname{\mathbb{I}m}\{\cdot\}$ is the imaginary part of the complex point. Normalized to an average power of 1, the complex symbol $x_k$ is then sent through the channel $c(\cdot)$. It should be emphasized that channel $c(\cdot)$ can be any channel for which an output can be generated with a given input. The output of the channel, denoted by $y_k$ is then applied to the decoder $NN_{d}(\mathbf{w}_d)$ which uses a softmax output layer to output a vector of posterior probabilities $\mathbf{s}_k \in \mathbb{R}^M$. The full data propagation throughout the AE is represented with $\mathbf{h}(\cdot)$ such that $\mathbf{s}_k = \mathbf{h}(k,\mathbf{w},\mathbf{u}_k)$.

The described AE structure can be considered an $M$--class pattern-classification problem. For such a problem, it is appropriate to optimize the weights $\mathbf{w}$ by minimizing the cross--entropy cost function \cite{goodfellow2016deep}
\begin{equation}
    \begin{split}
    J_{CE}(k,\mathbf{w}) & = -\sum_{i=1}^{M} \mathbf{t}^{(i)}_k \log \mathbf{s}^{(i)}_k \\
    & = -\sum_{i=1}^{M} \mathbf{t}^{(i)}_k \log \mathbf{h}^{(i)}(k,\mathbf{w},\mathbf{u}_k) \text{,}
    \end{split}
    \label{eq:Xent}
\end{equation}
where $\mathbf{t}_k$ is the target AE output and $(i)$ denotes the $i$-th output node of the decoder neural network $NN_{d}$. Even though the target AE output is $\mathbf{t}_k = \mathbf{u}_k$, the notation $\mathbf{t}_k$ will be used in the rest of the paper for clarity. The cross--entropy cost function can be used to calculate an AE-based upper bound $\hat{H}_{\hat{p}}(X|Y)=\mathbb{E}[\hat{p}_{X|Y}(x|y)]=\sum_{k=1}^K J_{CE}(k,\mathbf{w})$ to the true conditional entropy $H_p(X|Y)$, where $K$ is the number of symbols transmitted. Replacing $\hat{H}_{q}(X|Y)$ in Eq. (\ref{eq:MI_lower}) with $\hat{H}_{\hat{p}}(X|Y)$ an AE-based lower bound on the MI is obtained. Therefore, minimizing the cross--entropy cost function is equivalent to maximizing the MI between transmitted and received symbols, therefore satisfying the overall goal of the system. The weights $\mathbf{w}$ can be optimized by applying Bayesian filtering techniques \cite{arasaratnam2008nonlinear} as discussed in the following.
\section{Training of autoencoders using Bayesian filtering} \label{E2E-AE}

In order to apply Bayesian filtering techniques for the optimization, the AE structure depicted in Fig.~\ref{fig:AE} has to be described using a state--space modelling framework. The state--space model is described by a pair of equations, known as the \emph{process} equation and \emph{measurement} equation. The process equation describes the evolution of the states which are non--observable variables that we would like to estimate ($\mathbf{w}$ for the considered case). The corresponding measurement equation relates observable variables ($\mathbf{s}_k$ for the considered case) to the states \cite{arasaratnam2008nonlinear}.

In order to correctly perform the optimization of the AE using Bayesian filtering techniques, the state--space model has to be defined specifically for the system under consideration. In the following subsection \ref{space}, the space--state model will be constructed so that it fulfills the requirements of the AE structure and the desired performance metric (MI for the considered case). First, the process equation will be defined, followed by an adaptation of the measurement equation to fit the system requirements (cross--entropy cost function and batch optimization). In subsection \ref{CKF}, the optimization procedure will be detailed based on the established state--space model.

\subsection{Autoencoder state--space model} \label{space}

For the AE, the weights associated with the encoder and decoder NNs form column vectors constructed by stacking the weights associated with each neuron, starting with the first neuron in the first hidden layer and progressing first in width (next neuron) and then in depth (next layer). The two column vectors are then stacked to form a single column weight vector $\mathbf{w}_j=[\mathbf{w}^T_{e,j},\mathbf{w}^T_{d,j}]^T$, where $j$ is the $j$-th iteration step. The AE weights are considered as states and their evolution is described by the first--order auto-regressive equation \cite{arasaratnam2008nonlinear}:
\begin{equation}\label{eq:states}
\mathbf{w}_j = \mathbf{w}_{j-1} + \mathbf{q}_{j-1}     
\end{equation}
where $\mathbf{q}_j$ represents process--noise term which corresponds to white Gaussian noise of zero mean and covariance matrix $\mathbf{Q_{j-1}}$. For simplicity reasons, the covariance matrix is defined as a diagonal matrix $\mathbf{Q_{j-1}}=Q_{j-1} \mathbf{I}$, where $\mathbf{I}$ is an identity matrix. The process-noise term is intentionally included to avoid the states being trapped in a local minimum during the initial training stage.

The measurement equation for the state--space model describing the AE is defined such that it assumes that the target output $\mathbf{t}_{j}$ is a noisy measurement of the actual AE output:
\begin{equation}
    \mathbf{t}_{j}=\mathbf{h}(j,\mathbf{w}_{j},\mathbf{u}_{j})+\mathbf{r}_{j}\text{,}
    \label{eq:M_eq}
\end{equation}
where $\mathbf{r}_{j}$ is the measurement--noise term which corresponds to white Gaussian noise of zero mean and covariance matrix $\mathbf{R}_j$. The term $\mathbf{u}_{j}$ is the input of the AE at iteration $j$. With the measurement equation in its current form, Bayesian filtering techniques implicitly minimize the sum of squared errors cost function \cite{arasaratnam2008nonlinear},
\begin{equation}
    J(j,\mathbf{w}_{j})=\sum_{i=1}^{M}(\mathbf{t}^{(i)}_{j}-\mathbf{h}^{(i)}(j,\mathbf{w}_{j},\mathbf{u}_{j}))^2\text{.}
\end{equation}
However, such a cost function does not comply with the goal of maximizing MI between the transmitted and received symbols. Therefore, the measurement equation needs to be adapted to a form where the weight estimation $\mathbf{w}_j$ is obtained by minimizing cross-entropy cost function. The adaptation is done following the steps explained in \cite{arasaratnam2008nonlinear}. First, the measurement equation (\ref{eq:M_eq}) can be rewritten as
\begin{equation}
    \mathbf{0} = (\mathbf{t}_{j}-\mathbf{h}(j,\mathbf{w}_{j},\mathbf{u}_{j})) + \mathbf{r}_{j}\text{,}
\end{equation}
where the measurement is forced to take a vector-value $\mathbf{0}$. Then, the vector-valued measurement equation can be reformulated as a scalar-valued equation
\begin{equation}
    0 =\sqrt{\sum_{i=1}^{M}(\mathbf{t}^{(i)}_{j}-\mathbf{h}^{(i)}(j,\mathbf{w}_{j},\mathbf{u}_{j}))^2} + r_{j} = \sqrt{J(j,\mathbf{w}_{j})}  + r_j \text{,}
\label{eq:adapt}
\end{equation}
where $r_j$ is now a single dimension zero mean Gaussian noise with a variance $R_j$.

The scalar reformulation of the measurement equation offers significant benefits: 1) Reduces the computational complexity of the Bayesian filter optimization; 2) Improves the numerical stability of the Bayesian filter \cite{arasaratnam2008nonlinear}; 3) The cost function is directly incorporated into the measurement equation. The new formulation of the measurement equation allows various cost functions to be fitted and used with Bayesian filtering techniques. The final step of the adaptation is just incorporating the required cross--entropy cost function from Eq. (\ref{eq:Xent}) into Eq. (\ref{eq:adapt})
\begin{equation}\label{eq:measurements}
\begin{split}
    \tilde{t}_j = 0 & = \sqrt{-\sum_{i=1}^{M} \mathbf{t}^{(i)}_j \log \mathbf{h}^{(i)}(j,\mathbf{w}_{j},\mathbf{u}_j)} + r_j \\
    & = \tilde{h}(j,\mathbf{w}_{j},\mathbf{u}_{j},\mathbf{t}_{j}) + r_{j} \text{.}
\end{split}
\end{equation}
The Bayesian filtering techniques typically perform a single weight-vector update on the basis of a single input-output data pair. In the case of the AE, due to the normalization of the constellation, it would be advantageous to coordinate the weight update on a batch of data. Batch optimization is referred to as multistream training in Kalman filter related work and it is described in \cite{Multistream}. Multiple instances of the input, a batch of size $B$, are propagated through the system with the same weight set $\mathbf{w}_{j}$. The input to the AE is now $\mathbf{U}_j = [\mathbf{u}_{j \cdot B},\mathbf{u}_{j \cdot B+1},\dots,\mathbf{u}_{j \cdot B + (B-1)} ]^T$, each are propagated through the AE and form a vector-valued measurement,
\begin{equation}
    \begin{split}
    \mathbf{\tilde{t}}_j=\mathbf{0} & = [\tilde{t}_{j \cdot B},\tilde{t}_{j \cdot B+1},\dots,\tilde{t}_{j \cdot B + (B-1)} ]^T = \\
    & = \mathbf{\tilde{h}}(j, \mathbf{w}_{j}, \mathbf{U}_{j}, \mathbf{T}_j) + \mathbf{r}_{j}\text{,} \label{eq:Meas_batch_0}
    \end{split}
\end{equation}
where $\mathbf{T}_j$ is the target output corresponding to the AE input $\mathbf{U}_j$ and $\mathbf{r}_{j}=[r_{j \cdot B},r_{j \cdot B+1},\dots,r_{j \cdot B + (B-1)} ]^T$ is a vector that represents different independent noise realizations. A diagonal covariance matrix $\mathbf{R}_j = R_j \mathbf{I} \in \mathbb{R}^{B \times B}$ is used to describe $\mathbf{r}_{j}$.

Now that both process and measurement equations are defined to fulfill the requirements of the AE, a Bayesian filtering technique can be used to estimate the weights. Based on the Bayesian filtering paradigm, a complete statistical description of the state at iteration $j$ is provided by the posterior density of the state. Exploiting the new measurement, the old posterior density of the state is updated in two steps, \textit{prediction} and \textit{correction}. In the correction step, the posterior density of the state is computed exploiting the predictive density obtained in the prediction step. The weight estimation within the state--space model given by Eq. (\ref{eq:states}) and (\ref{eq:measurements}) can be solved by using various nonlinear Bayesian state estimation techniques \cite{arasaratnam2008nonlinear}. In this paper, the focus was put on using cubature Kalman filter (CKF) \cite{haykin2009cubature}. The advantage of using CKF is its accuracy and most importantly that it does not require computations of gradients providing a gradient--free training of AEs.

\subsection{Cubature Kalman filter} \label{CKF}

\begin{figure*}[!t] 
\centering
\includegraphics[width=\textwidth]{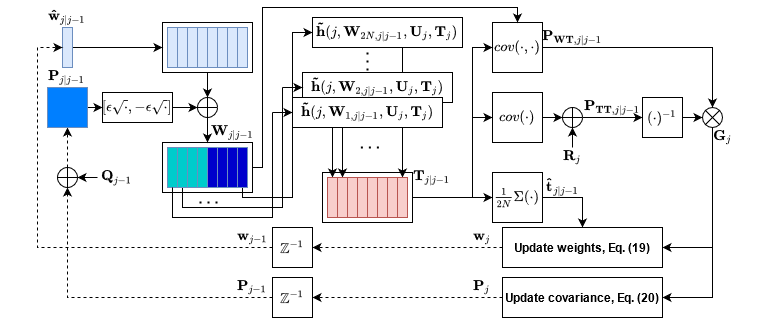}
\caption{Schematic of the optimization process. The covariance matrix calculation is denoted as $cov(\cdot)$, whereas $cov(\cdot,\cdot)$ denotes the cross-covariance matrix calculation. A bar is used to represent a column vector, a square to represent a square matrix and multiple bars encapsulated a rectangular matrix. Two colors have been used with $\mathbf{W}_{j|j-1}$ to indicated that one half of the matrix is a a result of an addition and the other the result of a subtraction. The unit time delay operator is denoted as $\mathbb{Z}^{-1}$.}
\label{fig:CKF}
\end{figure*}

The assumption of CKF is that the predictive and the posterior densities of the state are Gaussian distributions described by their means and covariances. In the prediction step, the predictive density of the current process state is computed based on the posterior density of the previous process state. The mean and covariance of the predictive density of the process state are denoted as $\mathbf{\hat{w}}_{j|j-1} \in \mathbb{R}^{N}$ and $\mathbf{P}_{j|j-1} \in \mathbb{R}^{N \times N}$, respectively. Since the process equation (\ref{eq:states}) yields a linear transition, $\mathbf{\hat{w}}_{j|j-1}$ and $\mathbf{P}_{j|j-1}$ are defined as:
\begin{gather}
\mathbf{\hat{w}}_{j|j-1}=\mathbf{w}_{j-1}, \label{eq:pred_w} \\
\mathbf{P}_{j|j-1}=\mathbf{P}_{j-1} + \mathbf{Q}_{j-1}. \label{eq:pred_P}
\end{gather}

In the correction step, the posterior density of the state is obtained from the predicted measurement density. In order to calculate the predicted measurement density, $2N$ cubature points are formed
\begin{equation}
    \mathbf{W}_{j|j-1}=[\mathbf{\hat{w}}_{j|j-1}+\epsilon\sqrt{\mathbf{P}_{j|j-1}} \quad \mathbf{\hat{w}}_{j|j-1}-\epsilon\sqrt{\mathbf{P}_{j|j-1}}] \in \mathbb{R}^{N \times 2N}\text{,}
    \label{eq:W}
\end{equation}
where $\sqrt(\cdot)$ is the square root of a matrix satisfying the relation $\mathbf{P}_{j|j-1}=\sqrt{\mathbf{P}_{j|j-1}}\sqrt{\mathbf{P}_{j|j-1}}^T$, and the unit cubature point $\epsilon = \sqrt{N}$.
Each column of the cubature points $\mathbf{W}_{i,j|j-1}$, where $i=1,2,\dots,2N$ represents the $i$-th column of $\mathbf{W}_{j|j-1}$, is a different realizations of the AE weights and the input is propagated through each of them to obtain the output of each realization $\mathbf{T}_{j|j-1} \in \mathbb{R}^{B \times 2N}$. The obtained outputs are used to calculate the predicted measurement $\mathbf{\hat{t}}_{j|j-1} \in \mathbb{R}^{B}$:
\begin{gather}
\mathbf{T}_{j|j-1}=\mathbf{\tilde{h}}(j, \mathbf{W}_{j|j-1}, \mathbf{U}_{j}, \mathbf{T}_j) \label{eq:h}\\
\mathbf{\hat{t}}_{j|j-1}=\frac{1}{2N}\sum_{i=1}^{2N}\mathbf{T}_{i,j|j-1}
\end{gather}
where $\mathbf{T}_{i,j|j-1}$ represents the $i$-th column of $\mathbf{T}_{j|j-1}$.
The covariance associated with the predicted measurement density $\mathbf{P}_{\mathbf{TT},j|j-1} \in \mathbb{R}^{B \times B}$, also known as the \emph{innovation covariance}, is estimated
\begin{equation}
\begin{split}
& \mathbf{P}_{\mathbf{TT},j|j-1} = cov(\mathbf{T}_{j|j-1}) + \mathbf{R}_{j} \\
& = \frac{1}{2N}\sum_{i=1}^{2N}(\mathbf{T}_{i,j|j-1}-\mathbf{\hat{t}}_{j|j-1})(\mathbf{T}_{i,j|j-1}-\mathbf{\hat{t}}_{j|j-1})^T+\mathbf{R}_{j},
\end{split}
\end{equation}
where $cov(\cdot)$ denotes the covariance matrix calculation. The cross--covariance $\mathbf{P}_{\mathbf{WT},j|j-1} \in \mathbb{R}^{N \times B}$ of the state and the measurement is calculated
\begin{equation}
\begin{split}
& \mathbf{P}_{\mathbf{WT},j|j-1} = cov(\mathbf{W}_{j|j-1},\mathbf{T}_{j|j-1}) \\
& =\frac{1}{2N}\sum_{i=1}^{2N}(\mathbf{W}_{i,j|j-1}-\mathbf{\hat{w}}_{j|j-1})(\mathbf{T}_{i,j|j-1}-\mathbf{\hat{t}}_{j|j-1})^T ,
\end{split}
\end{equation}
where $cov(\cdot,\cdot)$ denotes the cross--covariance matrix calculation. The Kalman gain $\mathbf{G}_j \in \mathbb{R}^{N \times B}$ is then calculated
\begin{equation}
\mathbf{G}_j=\mathbf{P}_{\mathbf{WT},j|j-1}\mathbf{P}_{\mathbf{TT},j|j-1}^{-1}, \label{eq:gain}   
\end{equation}
and used to update the weights $\mathbf{w}_j$ conditional on the measurement
\begin{equation}
\begin{split}
\mathbf{w}_j & =\mathbf{\hat{w}}_{j|j-1}+\mathbf{G}_j(\mathbf{\tilde{t}}_j-\mathbf{\hat{t}}_{j|j-1}) \\
& =\mathbf{\hat{w}}_{j|j-1}+\mathbf{G}_j(-\mathbf{\hat{t}}_{j|j-1})\label{eq:update}
\end{split}
\end{equation}
Based on Eq. (\ref{eq:Meas_batch_0}), the target output is $\mathbf{\tilde{t}}_j=\mathbf{0}$, therefore it can be neglected in Eq. (\ref{eq:update}). The Kalman gain $\mathbf{G}_j$ is also used to update the covariance $\mathbf{P}_{j}$
\begin{equation}
\mathbf{P}_{j}=\mathbf{P}_{j|j-1}-\mathbf{G}_j\mathbf{P}_{\mathbf{TT},j|j-1}\mathbf{G}_j^T.
\label{eq:P_update}
\end{equation}

Observing Eq. (\ref{eq:W}--\ref{eq:gain}), it can be noticed that the complexity of the algorithm depends on the number of weights $N$ and batch size $B$. The noise covariance matrices $\mathbf{Q_{j-1}}$ and $\mathbf{R_j}$ are hyperparameters of the CKF algorithm and its performance depends on how they are chosen. Fig. \ref{fig:CKF} illustrates the optimization process described by Eq. (\ref{eq:pred_w}--\ref{eq:P_update}).

\section{Communication system description} \label{system}

\subsection{AE model}

In \cite{Gumus2020}, it was demonstrated that state-of-the-art generalized mutual information (GMI) performance can be achieved with an encoder NN with no hidden layers and with biases set to zero. Since GMI is a good indicator of system performance \cite{Alvarado:15}, an encoder NN with no hidden layers and with biases set to zero has been used in this paper. As a result of a coarse optimization, the decoder neural network has a single hidden layer of $\frac{M}{2}$ nodes and \emph{Leaky Relu} as the activation function. The AE architecture is summarized in Table \ref{tb:Ae_arch}. The initial weight vector $\mathbf{\hat{w}}_0$ is initialized using \textit{Glorot initialization} \cite{Glorot2010}, whereas the initial covariance $\mathbf{P}_{0}$ is an identity matrix.

\begin{table}[!t]
\renewcommand{\arraystretch}{1.3}
\caption{Parameters of the encoder and decoder neural network}
\label{tb:Ae_arch}
\centering
\begin{tabular}{|c||c|c|}
\hline
 & Encoder NN & Decoder NN\\
\hline
\# of input nodes & M & 2\\
\hline
\# of hidden layers & 0 & 1\\
\hline
\# of nodes per hidden layer & 0 & $M/2$\\
\hline
\# of output nodes & 2 & M\\
\hline
Bias & No & Yes\\
\hline
Hidden layer activation function & None & Leaky Relu\\
\hline
Output layer activation function & Linear & Softmax\\
\hline
\end{tabular}
\end{table}

\subsection{Embedded channel models}
\begin{figure}[!t]
\centering
\includegraphics[width=\columnwidth]{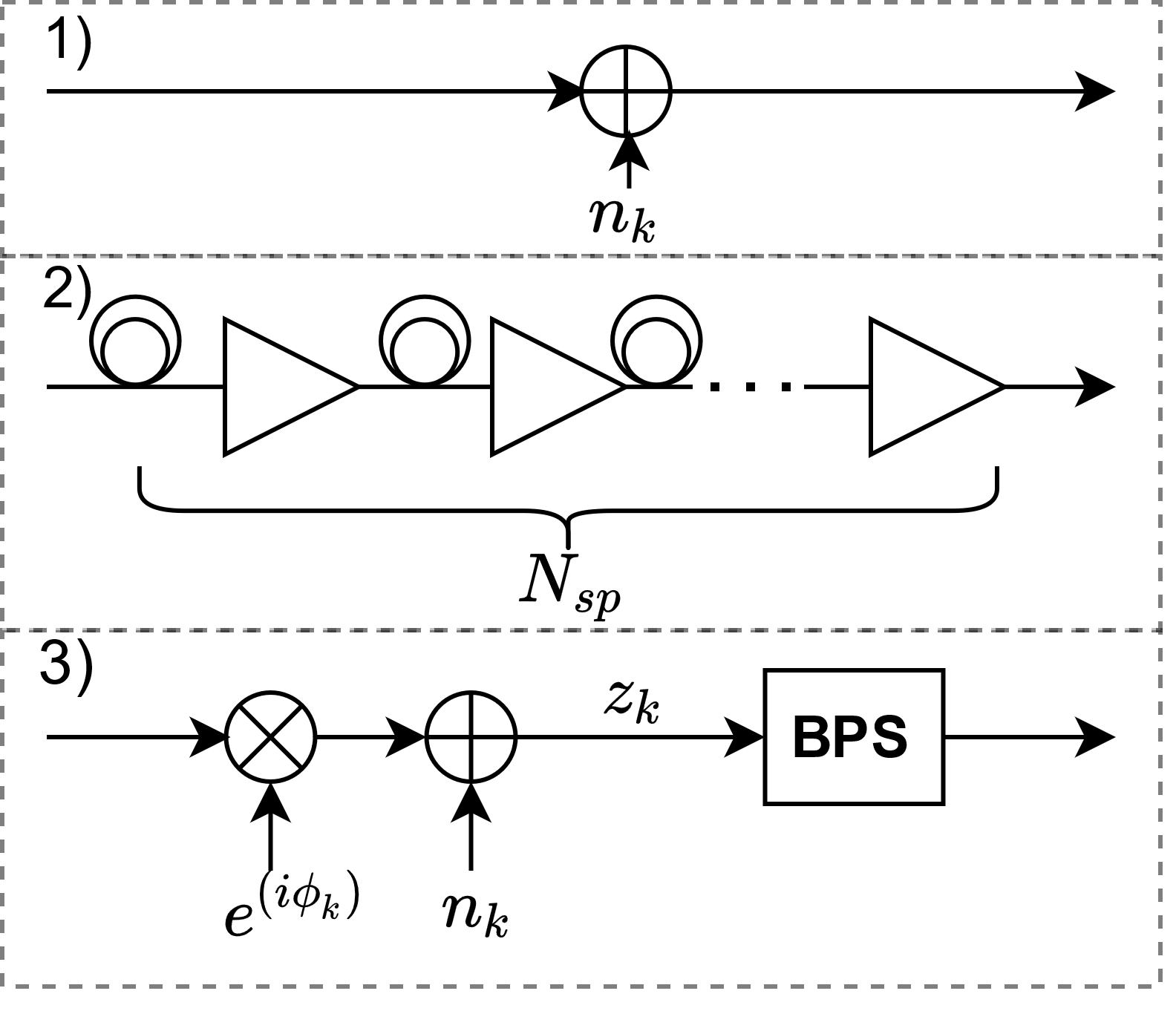}
\caption{Embedded channel models consist of: 1) additive white Gaussian noise (AWGN); 2) nonlinear phase noise (NLPN) 3) phase noise, AWGN and blind phase search algorithm for phase recovery. }
\label{fig:channels}
\end{figure}
In this paper, three channel models that operate on one sample per symbol basis, with symbol rate $R_s$, are embedded into the autoencoder. Those channel models are illustrated in Fig. \ref{fig:channels}:
\subsubsection{AWGN channel}

The noise variance is determined by the signal-to-noise ratio (SNR): $\sigma_N^2=\frac{1}{SNR}$.
As the AWGN channel model is differentiable, a fair comparison in the performance of the AE trained using the CKF and the gradient--based methods can be obtained. For the considered case, we employ the backpropagation algorithm using the Adam optimizer as a benchmark gradient--based training method \cite{kingma2014adam}.

\subsubsection{Nonlinear phase noise channel}
A memoryless fiber channel model that includes multiple spans $N_{sp}$ with ideal lumped amplification is considered. Erbium-Doped Fiber Amplifiers (EDFAs), described by the noise figure $NF$ and amplifier gain $G$, are used and they introduce \emph{amplified spontaneous emission} (ASE) noise $n_{ASE}$. Similar to \cite{Li2018b}, the channel model is obtained from the nonlinear Schr\"{o}dinger equation by neglecting the dispersion. The channel model shows the impact of a data-dependent nonlinear phase shift known as the \emph{nonlinear phase noise} (NLPN) and it is defined by the recursion
\begin{equation}
z_{k}^{(i+1)} = z_{k}^{(i)} e^{i\gamma L_{eff} |z_{k}^{(i)}|^2} + n_{ASE,k}^{(i+1)}, \quad 0 \leq i < N_{sp},
\end{equation}
where $(i)$ is $i$-th fiber span, $\gamma$ is the nonlinearity parameter, and $L_{eff}=(1-e^{\alpha L})/\alpha$ is the effective length of a span. The actual length of a span is $L$ and $\alpha$ is the attenuation coefficient. The noise term $n^{i+1}_{ASE}$ is a zero--mean Gaussian distribution with variance $P_n= h F_c R_s (G \cdot NF-1)/2$ \cite{Essiambre2010}, where $h$ is Planck's constant and $F_c$ is the carrier frequency. The input to the channel is a power rescaled signal $z_k^{(0)}=\sqrt{P_{in}}x_k$, where $P_{in}$ the launch power. The output of the channel $y_k = z_k^{(N_{sp})}/P_{in}$ is normalized before processed by the decoder.

The NLPN channel model serves as the second differentiable channel model to use for comparison of the CKF and gradient--based methods. As for the previous channel model, the Adam optimizer is used as a benchmark gradient--based method.

\subsubsection{Non--differentiable phase noise channel}
As a validation of our proposed optimization method, we consider a channel model for which the gradients cannot be computed. For this purpose, the phase noise channel with blind phase search (BPS) \cite{Pfau2009a}, which is a standard phase noise compensation algorithm, is considered. 

The encoder NN output $x_k$ is distorted by phase noise and additive noise
\begin{equation}
    z_k = x_k e^{i \phi_k} + n_k \text{,}
\end{equation}
where the AWGN term $n_k$ is characterized by noise power $\sigma^2_N$. The phase noise $\phi_k$ is modeled as a Wiener process
\begin{equation}
    \phi_k = \phi_{k-1} + \Delta \phi_k \text{,}
\end{equation}
where $\Delta \phi_k$ is the random phase increment with a zero-mean and variance  $\sigma^2_\phi = 2\pi\Delta\nu T_s$. The combined transmitter and receiver oscillator linewidth is denoted by $\Delta \nu$, and $T_s$ is the symbol period.

The BPS is used as the phase recovery algorithm and it is the non-differentiable part of this channel model due to its hard-decision directed nature \cite{Pfau2009a}. The BPS is a pure feedforward phase recovery algorithm based on rotating the received symbol by $N_s$ test phases defined by:
\begin{equation}
    \theta_i=\frac{i}{N_s}\cdot 2\pi, \quad i \in \{0,1,\dots,N_s-1 \},
\end{equation}
where $i$ represents the $i$-th test phase. Decisions are made on each of the rotated symbols and the distance between the decided symbol $\hat{z}_{k,i}$ and the rotated symbol $z_{k,i}$ is calculated. Afterwards, distances of symbols rotated by the same test phase are summed over a window of size $W_{BPS}$
\begin{equation}
    d_{k,i}=\sum_{j=1}^{W_{BPS}}|z_{k,i}^{(j)}-\hat{z}_{k,i}^{(j)}|^2,
\end{equation}
where $(j)$ is the $j$-th sample in the window. This step mitigates the effect of the AWGN on the quality of the decision. Finally, the optimal test phase is chosen by the minimum sum of distances \cite{Pfau2009a}
\begin{equation}
    \hat{\phi}_k=\underset{\theta_i}{\mathrm{argmin}} \quad d_{k,i},
\end{equation}
where $\mathrm{argmin}$ is a non--differentiable operation.
The received symbol is rotated by the chosen test phase to output the phase compensated sample
\begin{equation}
    y_k=z_k e^{-i\hat{\phi}_k}.
\end{equation}
The performance of the BPS algorithm is determined by the parameters $N_s$ and $W_{BPS}$.
\subsection{Gaussian receiver} 
For optical communication, it is a common approach to use a mismatched Gaussian receiver \cite{Lapidoth1994,Metodi:16} and
assume the transition probability $q_{Y|X}(y|x)$ in Eq. (\ref{eq:MI_lower}) is of an auxiliary Gaussian channel
\begin{equation}
    q_{Y|X}(y|x) = \frac{1}{\sqrt{2 \pi \sigma_{G}^2}}\exp{(-\frac{||y-x||^2}{2 \sigma_{G}^2})} \text{,}
\end{equation}
where $\sigma_{G}^2$ is the noise variance and $||y-x||^2$ is a squared Euclidean distance between the symbols.
Applying the Bayes' theorem, the conditional probability that a specific $x$ was sent observing $y$ is given as
\begin{equation}
    q_{X|Y}(x|y) = \frac{p_X(x)q_{Y|X}(y|x)}{\sum_{x^i \in \mathcal{X}} p_X(x=x^i)q_{Y|X}(y|x=x^i)} \text{.}
    \label{eq:qxy}
\end{equation}
The Monte Carlo approach can be used to evaluate Eq. (\ref{eq:qxy}). The noise variance $\sigma_{G}^2$ can be estimated from the channel input-output pairs.

In the case of the AWGN channel model, the assumed transition probability $q_{Y|X}(y|x)$ is identical to the true $p_{Y|X}(y|x)$. Therefore, the Gaussian receiver is the maximum likelihood (ML) receiver for the AWGN channel and the estimated noise variance $\sigma_{G}^2 = \sigma_{N}^2$. Therefore, on the AWGN channel, the performance of the Gaussian receiver is an upper bound for the performance of the NN decoder.

However, in the case of channel models 2) and 3), the true transition probability $p_{Y|X}(y|x)$ is approximated by $q_{Y|X}(y|x)$. The combined distortion of the nonlinear phase noise and the additive noise of channel model 2) is assumed to be purely Gaussian and approximated with the noise variance $\sigma_{G}^2$. The same assumption and approximation is made for the combined distortion of the residual phase noise and the additive noise of channel model 3). Therefore, the Gaussian receiver is a mismatched receiver for both of these channel models.

\section{Numerical results} \label{results}

The system symbol rate is $R_s=32$~GBd and the size of the constellation is $M=64$. At each training iteration $j$, a batch $\mathbf{U}_j$ of $B=32\cdot M$ one-hot encoded input vectors $\mathbf{u}_k$ is generated. The hyperparameters of the CKF algorithm $Q_{j-1}$ and $R_j$ are coarsely optimized using a standard grid search method. Both hyperparameters are sampled from the set $\{1,10^{-1},10^{-2},10^{-3},10^{-4},10^{-5},10^{-6} \}$ and the CKF algorithm is applied with each combination. The combination that achieves the best mutual information is chosen. The training was done for each SNR value. The AE is trained until the cost function converges. Afterwards, the weights of the AE are fixed and testing is performed. The testing was done by running $100$ simulations with $10^5$ symbols per simulation. Each of the trained AE was tested with the same channel parameters as it was trained.

The constellations learned by AE will be compared to iterative polar modulation (IPM) \cite{Djordjevic2010} and square QAM (referred to simply as QAM in the following). The IPM was chosen as a benchmark because it is a near-optimal constellation shape for the AWGN channel \cite{Djordjevic2010}.

\subsection{AWGN channel}
 
The studied SNR region includes values from the interval SNR $=\{10,11,\dots,25\}$ dB. Since the Gaussian receiver is optimal for the AWGN channel, the presented results for this channel model only include the Gaussian receiver.

\begin{figure}[!t]
\centering
\includegraphics[width=\columnwidth]{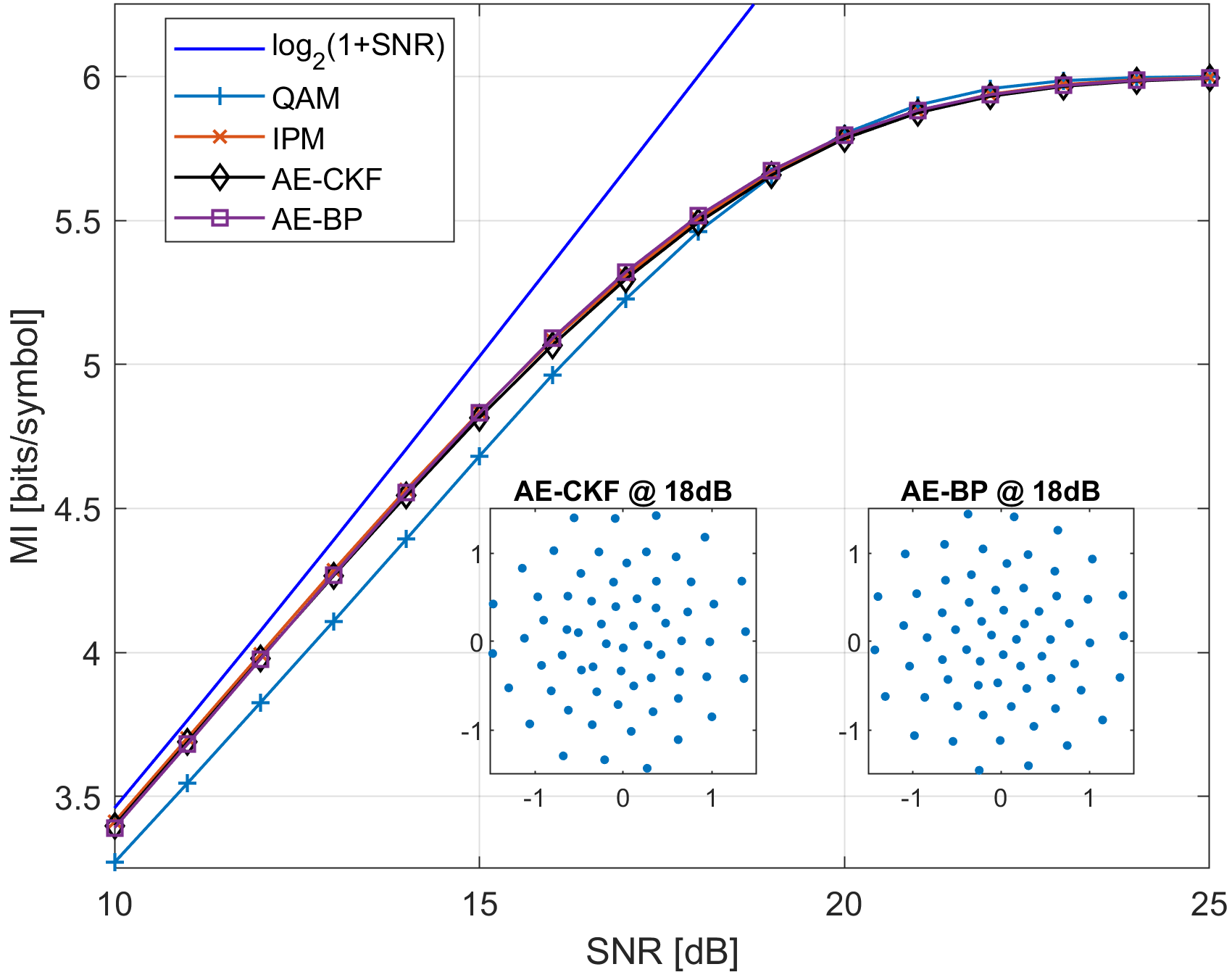}
\caption{\emph{AWGN channel:} Mutual information as a function of SNR for constellation size $M=64$. Inset: example of the learned constellations with CKF and BP at SNR $=18$~dB.}
\label{fig:AWGN}
\end{figure}

The simulation results of the testing for the AWGN channel are shown in Fig. \ref{fig:AWGN}, which illustrates the performance in MI with respect to the SNR. The learned constellations using the AE trained with the CKF and the backpropagation are denoted as AE-CKF and AE-BP, respectively. This notation will be used throughout the rest of this section. The learned constellations AE-CKF and AE-BP result in a similar performance in the MI with an average difference of around $0.01$ bits/symbol. Compared to QAM, the two constellations achieve higher MI in the low SNR region, as expected, whereas the difference is marginal compared to the IPM. 

The insets illustrate the learned constellations AE-CKF and AE-BP when training on SNR $=18$~dB. The Euclidian distance between some of the points in AE-CKF is quite small, but this does not have a negative effect on the MI performance. Maximum shaping gain is not achieved by maximizing Euclidean distance but rather by optimizing the trade-off between the Euclidian distance and the energy distribution of the signal. As expected for an AWGN channel, the optimizer pushes points towards the origin in order to resemble the AWGN-optimal Gaussian distribution of the signal energy, thus providing overall gain.

\subsection{NLPN channel}
The training of the AE is performed by sweeping the launch power $P_{in}$ in the interval $P_{in}=\{-8,-7.5,\dots,0\}$~dBm for the fiber parameters $\gamma=1.27 \frac{1}{W \cdot km}$, $\alpha=0.2 \frac{\text{dB}}{km}$, $NF=5$~dB, $F_c=193.41$~THz, $N_{sp}=10$, and $L=100$~km. The MI estimation for the QAM and IPM constellation is performed using the mismatched Gaussian receiver. The results for the constellations learn by applying CKF (AE-CKF) and backpropagation (AE-BP) include the MI estimation for both mismatched Gaussian receiver and decoder NN.

\begin{figure}[!t]
\centering
\includegraphics[width=\columnwidth]{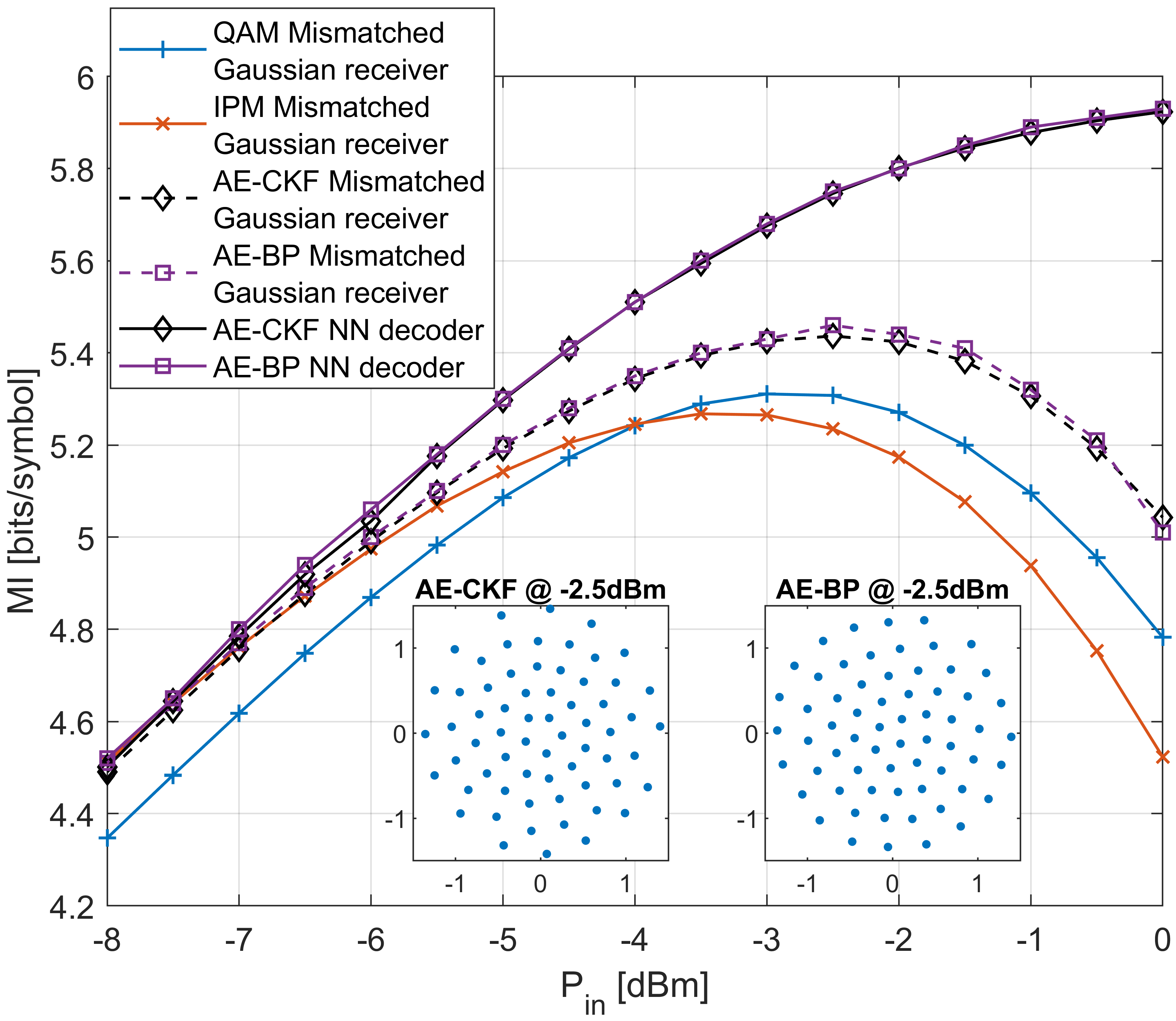}
\caption{\emph{NLPN channel:} Mutual information as a function of launch power $P_{in}$ for constellation size $M=64$. Inset: example of the learned constellations with CKF and BP for launch power $P_{in}=-2.5$~dBm.}
\label{fig:NLPN}
\end{figure}

In Fig. \ref{fig:NLPN}, the MI performance with respect to launch power $P_{in}$ is shown. For AE-CKF and AE-BP, the dashed lines represent the MI performance when the mismatched Gaussian receiver is used and the solid lines represent the MI performance when using the decoder NN. The learned constellations AE-CKF and AE-BP outperform both QAM and IPM in terms of MI, with both mismatched Gaussian receiver and decoder NN. Observing just the mismatched Gaussian receiver results, the learned constellations achieve a greater maximum MI by up to $0.12$ and $0.17$~bits/symbol compared to QAM and IPM, respectively. Observing the studied launch power $P_{in}$ region, the maximum gain in MI compared to QAM was obtained at $P_{in}=0$~dBm and it amounts to around $0.26$~bits/symbol. At the same launch power, the maximum gain in MI compared to IPM was achieved, reaching a value of up to $0.52$~bits/symbol.  It can be concluded that the learned constellations are more robust to nonlinear phase noise than QAM and IPM.

The two learned constellations AE-CKF and AE-BP have similar performance for both receivers in question. The average difference for both the mismatched Gaussian receiver and the decoder NN is around $0.01$~bits/symbol. Observing the MI performance obtained with the decoder NN, similar mitigation and compensation was observed as what was shown in \cite{Li2018b}. The insets illustrate the learned constellations AE-CKF and AE-BP when training on launch power $P_{in}=-2.5$~dBm.

The obtained results demonstrate that the proposed AE optimization method, CKF, can achieve similar performance to BP when the derivative of the cost function with respect to the encoder weights $\mathbf{w_e}$ can be calculated.

\subsection{Non--differentiable phase noise channel}

The training of AE is performed for each of the SNR values in the interval SNR $=\{15,16,\dots,20\}$ dB combined with the BPS parameters $N_s=36$ and $W_{BPS}=\{40,64\}$ and linewidth $\Delta\nu=100$~kHz. A mismatched Gaussian receiver is used instead of the decoder NN for the MI estimation during the testing phase in order to have a fair comparison between all the constellations.

\begin{figure}[!t]
\centering
\subfloat[]{\includegraphics[width=\columnwidth]{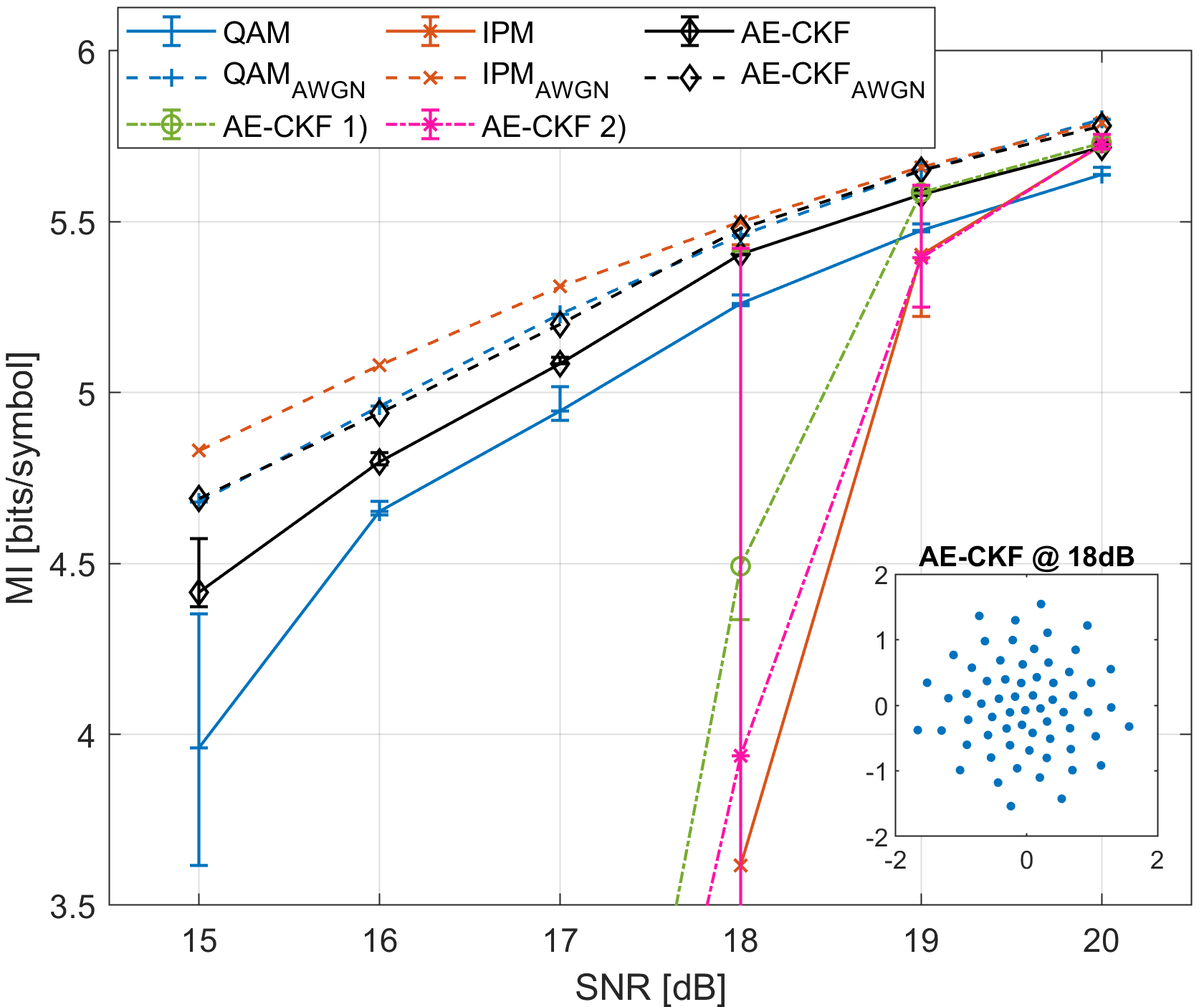}}
\hfil
\subfloat[]{\includegraphics[width=\columnwidth]{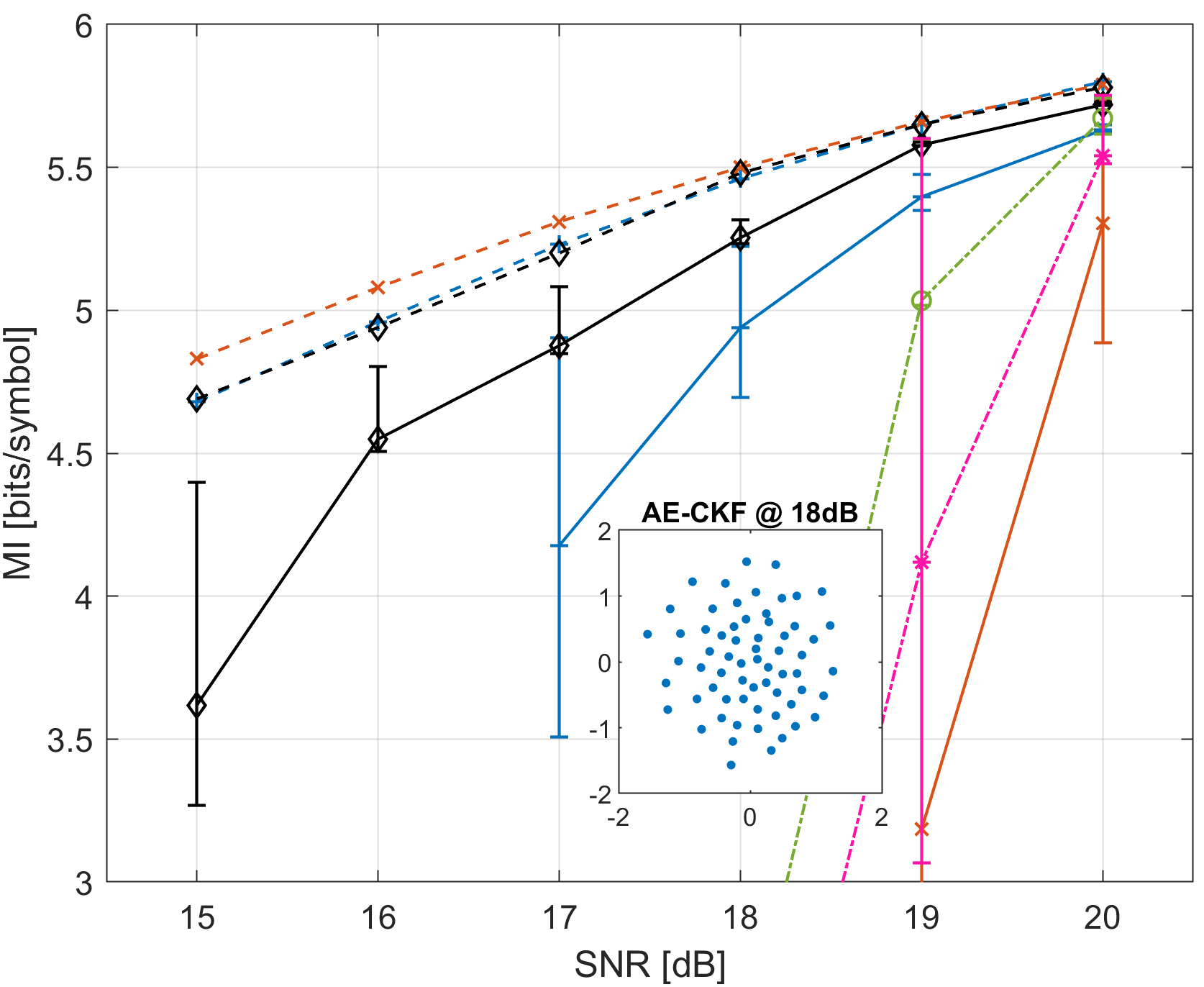}}
\caption{\emph{Nondifferentiable phase noise channel:} Mutual information as a function of SNR for constellation size M=64. The parameters of the BPS are: (a) $N_s=36$ and $W_{BPS}=64$ and (b) $N_s=36$ and $W_{BPS}=40$. The line shows the mean value, whereas the upper limit of the error bar is the max value and the lower limit is the 25th percentile. Points that have a mean less than 3.5 and 3 bits/symbol are omitted from the figure (a) and (b) for visual clarity, respectively. Inset: example of the learned constellation at SNR $=18$ dB.}
\label{fig:BPS}
\end{figure}

In Fig. \ref{fig:BPS}(a)-(b), the MI performance with respect to SNR is shown. The dashed lines show the performance of the constellations for the AWGN channel. The dashed line $\text{AE-CKF}_{AWGN}$ presents the MI obtained when the AE-CKF trained for channel model 3) is tested on channel model 1) for the respected SNR. These results are added in order to observe the penalty introduced by the phase noise. The solid lines represent the average MI value over the $100$ test simulations at a given SNR, whereas the upper limit of the error bar is the maximum obtained MI value and the lower limit shows the 25th percentile. Therefore, the error bars represent the $75\%$ of simulations with the highest MI. Constellations optimized on channel models 1) and 2) have been tested on channel model 3) and the curves representing the obtained results are denoted as "AE-CKF 1)" and "AE-CKF 2)", respectively. The insets illustrate the learned constellations AE-CKF when training on SNR$=18$~dB.

Fig. \ref{fig:BPS}(a) shows MI as a function of SNR when the BPS parameters are $N_s=36$ and $W_{BPS}=64$. For the SNR values with a mean MI less than $3.5$ bits/symbol, all points are removed from the figure for visual clarity. For the used BPS parameters, phase slips occur at SNR $=15$~dB for QAM constellation resulting in a wide error bar. The IPM constellation is designed for the AWGN channel and it is not suitable for a phase noise channel. The BPS algorithm does not estimate the phase accurately resulting into wide error bars. Observing the studied SNR region, the AE learned constellation achieves the highest gain in MI compared to QAM at SNR $=15$~dB, and it amounts to around $0.45$ bits/symbol. The gain in MI compared to QAM is $0.15$ bits/symbol at SNR $=16$ dB and with the increase of SNR the gain is slowly decaying to $0.08$ bits/symbol, which is achieved at SNR $=20$ dB. Observing the width of the error bars, it can be noticed that the spread of MI performance is smaller for the AE learned constellation compared to QAM, especially for lower SNR values. It can also be noticed that the maximum achieved MI value of QAM is lower than the 25th percentile of the AE learned constellation, implying that the AE learned constellation has a better performance at least in $75\%$ of the test cases. The AE-CKF and IPM constellations achieve similar maximum MI at high SNR but the AE-CKF is more robust and has a significantly higher mean MI value, whereas at SNR $=20$ dB the two constellations have marginal differences in MI. The AE-CKF 1) achieves similar performance as AE-CKF for SNR~$=\{19,20\}$~dB due to the fact that the main impairment at these SNR values is AWGN. Even though both IPM and AE-CKF 1) are optimized for AWGN channel, the AE-CKF 1) performs better than IPM when applied to the non-differentiable channel. However, the AE-CKF 2) has similar performance as IPM. As expected both AE-CKF 1) and 2) do not perform well for the given non-differentiable channel since they were not optimized for it.

Next, the potential improvement at lower BPS complexity is analyzed. The window size was decreased to $W_{BPS}=40$ and the results are illustrated in Fig. \ref{fig:BPS}(b). All of the SNR points with a mean MI less than $3$ bits/symbol were removed from the figure for visual clarity. Reducing the complexity of the BPS affects its performance, resulting in less accurate phase estimations. Therefore, the mentioned issues regarding QAM and IPM occur even for higher SNR values. The gain in MI achieved by the learned constellation compared to QAM at SNR $=20$ dB is similar to the one achieved in the previous scenario, whereas compared to IPM for the same SNR $=20$ dB greater gain was achieved, reaching a value of around 0.41 bits/symbol. The achieved gain is increasing with the lowering of the SNR. The gain at SNR $=17$ dB already exceeds the highest gain achieved in the previous scenario, reaching a gain of around $0.7$ bits/symbol. Similarly to the previous scenario, the error bars of the AE-CKF constellation are less spread than for QAM and IPM at the same SNR, indicating better robustness of the constellation. The less accurate phase estimation affects the performance of AE-CKF 1) and AE-CKF 2) which are in this scenario more penalized than QAM, but still less than IPM. At SNR~$=20$~dB, the AE-CKF 1) stands out with a higher MI than QAM, but still lower than AE-CKF.

The dashed lines in Fig. \ref{fig:BPS} show that the AE-CKF constellation learned by training on channel model 3) lost AWGN shaping gain compared to when it was trained directly on the AWGN channel. The $\text{AE-CKF}_{AWGN}$ and $\text{QAM}_{AWGN}$ now have similar performance in MI. Comparing situations with and without phase noise, the penalty in MI is significantly lower for AE-CKF than for QAM. This demonstrates that the AE-CKF constellation is more robust to phase noise at an expense of slight loss of AWGN shaping gain. 

\begin{figure}[!t]
\centering
\includegraphics[width=0.95\columnwidth]{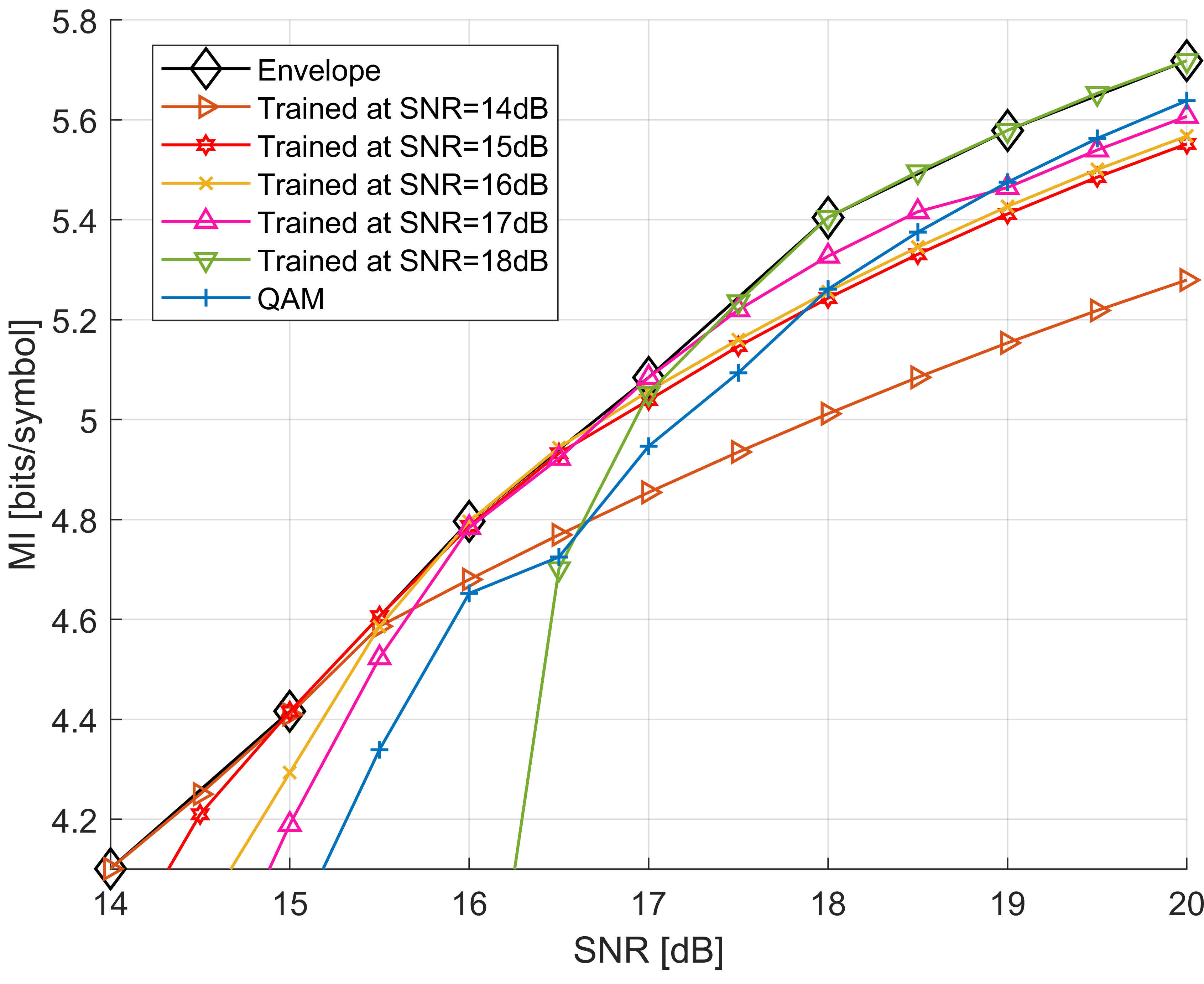}
\caption{\emph{Nondifferentiable phase noise channel:} Mutual information as a function of SNR for constellation size $M=64$ and BPS parameters $N_s=36$ and $W_{BPS}=64$. Sweeping the observed SNR interval with constellations trained with different SNR values.}
\label{fig:Envelope}
\end{figure}

The previous figures show results when the AE was trained and tested on the same SNR value to show the best achieved MI performance at the given SNR. However, in practice an SNR estimation error of up to $\sim 1$~dB can be encountered \cite{khan2016optical}. Fig. \ref{fig:Envelope} shows what the MI performance of constellations is when they are tested on different SNR values than what they were trained on. The results are only for the case when the BPS parameters are $N_s=36$ and $W_{BPS}=64$ and the curve denoted as "Envelope" is the same as "AE-CKF" curve from Fig. \ref{fig:BPS}(a) and it indicates the training points as well. It has been expanded with the MI obtained when training at SNR~$=14$~dB, in order to show the robustness of the constellation trained at SNR~$=15$~dB. The results obtained when training on SNR~$=\{18,19,20\}$~dB are similar and therefore only the results obtained for SNR~$=18$~dB are shown. All of the constellations show at least up to $\pm 0.5$~dB of robustness to SNR estimation errors with less than $1\%$ penalty in MI. The constellation trained at SNR~$=18$~dB can be used for the SNR~$=[17,20]$~dB interval with less than $1\%$ penalty in MI but at an expense of a considerable performance deterioration for SNR~$<17$~dB.

\begin{figure}[!t]
\centering
\includegraphics[width=0.95\columnwidth]{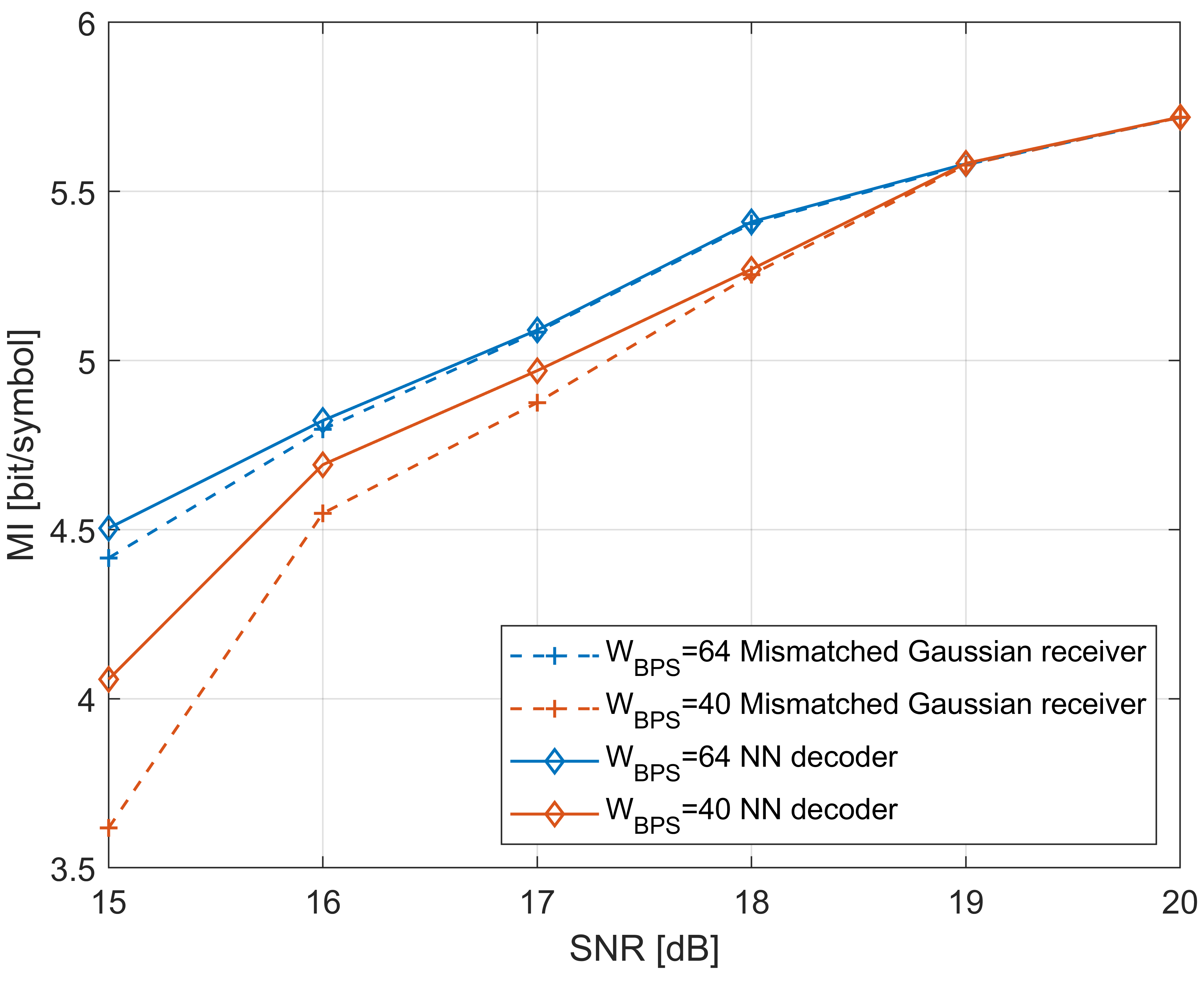}
\caption{\emph{Nondifferentiable phase noise channel:} Mutual information as a function of SNR for constellation size $M=64$. The dashed lines represent the mismatched Gaussian receiver, whereas the solid lines represent the NN receiver.}
\label{fig:BPS_decoder}
\end{figure}

It has been established that AE-CKF constellations outperform QAM and IPM when using the mismatched Gaussian receiver. Now, a comparison between the mismatched Gaussian receiver and NN decoder for AE-CKF constellations is provided, since the encoder and decoder come as a pair. In Fig. \ref{fig:BPS_decoder} the MI is illustrated for the mismatched Gaussian receiver with dashed lines and NN decoder with solid lines for $W_{BPS}=64$ and $W_{BPS}=40$. For lower SNR, when the estimation of the phase noise is worse, the NN decoder outperforms the Gaussian receiver. As the SNR increases the estimation of the phase noise improves, resulting in AWGN being the main source of distortion. Thus, the gain in MI that NN decoders achieve compared to the mismatched Gaussian receiver decreases since the mismatched Gaussian receiver is optimal for the AWGN channel. For SNR$=\{19,20\}$~dB the estimation of the phase noise is highly accurate for all the observed cases, therefore all have similar performance in regards to MI. In the case of $W_{BPS}=64$, the estimation of the phase noise is still quite accurate at SNR$=15$~dB and the NN decoder achieves around $0.09$~bits/symbol gain compared to the Gaussian receiver. Whereas, in the case of $W_{BPS}=40$, the estimation of the phase noise is significantly degraded at SNR$=15$~dB and the NN decoder achieves up to around $0.44$~bits/symbol gain compared to the Gaussian receiver.

\section{Discussion} \label{Discussion}
In this paper, the differentiable channel model requirement is lifted by proposing a derivative-free optimization method for AEs. The proposed method is exemplified by adopting CKF for the AE weights optimization. Most notably, the results of this study imply that the AE can be optimized with arbitrary black--box channels and not only with known differentiable ones. Although the training capabilities of the proposed method were demonstrated in few scenarios, the proposed method should allow the training of autoencoders on arbitrary channel models, e.g. experimental test--beds, which is the ultimate goal. This is left for future work.

The CKF algorithm can be applied for training of recurrent NNs, which is important if considering a channel with finite memory, such as the dispersive optical channel. The proposed state spaced framework supports training of recurrent NNs \cite{arasaratnam2008nonlinear}, which showed superior performance compared to feed-forward NNs when applied to a channel with finite memory \cite{Karanov2019a}. In such a case, the weight set is just expanded with the weights of the recurrent connections, meaning an identical state-based model for the kernel weights is used. The CKF derivation from this paper is therefore valid. This study is out of the scope of this paper, but is an interesting direction for future work.

\section{Conclusion} \label{Conclusion}
We have proposed and numerically demonstrated a derivative-free method for training autoencoders for geometrical constellation shaping. This is achieved by expressing the autoencoder weights and system as state-space models and then applying cubature Kalman filter (CKF) for state (encoder and decoder weights) estimation.  
For differentiable AWGN and NLPN channel models, it was shown that the performance, in terms of the mutual information of the learned constellations, is almost identical to when the training is performed using the standard backpropagation algorithm. The CKF trained autoencoder was also tested for a phase noise channel with a non-differentiable phase recovery algorithm. In such a case, the autoencoder-learned constellations achieved significant performance and robustness improvement with respect to conventional constellation shapes optimized for an AWGN channel. It should be emphasized that the proposed method can be applied to any AE structure and not just for GCS as it was used in this paper.

\section*{Acknowledgment}

This work was financially supported by the European Research Council through the ERC-CoG FRECOM project (grant agreement no. 771878), the Villum Young Investigator OPTIC-AI project (grant no. 29334), and DNRF SPOC, DNRF123.

\bibliographystyle{IEEEtran}
\bibliography{references.bib}
\end{document}